\documentclass[journal]{IEEEtran}
\usepackage{cite}
\usepackage{amsmath,amssymb,amsfonts}
\usepackage{algorithmic}
\usepackage{graphicx}
\usepackage{textcomp}
\usepackage{xcolor}
\usepackage{float}
\usepackage{xcolor}
\usepackage{siunitx}

\ifCLASSINFOpdf
\else
\fi
\begin{document}

\title{\emph{In situ} Performance of the Low Frequency Array for Advanced ACTPol}


\author{Yaqiong Li,
        Jason E. Austermann, 
        James A. Beall, 
        Sarah Marie Bruno,
        Steve K. Choi,
        Nicholas F. Cothard,
        Kevin T. Crowley,
        Shannon M. Duff,
        Shuay-Pwu Patty Ho,
        Joseph E. Golec,
        Gene C. Hilton,
        Matthew Hasselfield,
        Johannes Hubmayr,
        Brian J. Koopman,
        Marius Lungu,
        Jeff McMahon,
        Michael D. Niemack,
        Lyman A. Page,
        Maria Salatino,
        Sara M. Simon,
        Suzanne T. Staggs, 
        Jason R. Stevens,
        Joel N. Ullom,
        Eve M. Vavagiakis,
        Yuhan Wang,
        Edward J. Wollack,
        Zhilei Xu
\thanks{This work was supported by the U.S. National Science Foundation through award 1440226. The development of multichroic detectors and lenses was supported by NASA grants NNX13AE56G and NNX14AB58G. The work of KPC, KTC, EG, JEG, BJK, CM, BLS, JTW, and SMS was supported by NASA Space Technology Research Fellowship awards. SKC acknowledges support from the Cornell Presidential Postdoctoral Fellowship and the U.S. National Science Foundation award AST-2001866}
\thanks{Y. Li, S. K. Choi, N. F. Cothard, M. Niemack, E. M. Vavagiakis are with the Department of Physics and the Department of Astronomy, Cornell University, Ithaca, NY 14853, USA (e-mail: yl3549@cornell.edu; skc98@cornell.edu; nc467@cornell.edu; niemack@cornell.edu; ev66@cornell.edu)}
\thanks{J. Austermann, J. A. Beall, S. M. Duff, G. C. Hilton, J. Hubmayr, J.R. Stevens and J. N. Ullom are with NIST, Boulder, CO 80305, USA (e-mail: jason.austermann@nist.gov; james.beall@nist.gov; shannon.duff@nist.gov; gene.hilton@nist.gov; johannes.hubmayr@nist.gov; Jason.Stevens-2@colorado.edu; joel.ullom@nist.gov).}
\thanks{S. Bruno, M. Lungu, L. A. Page, S. Staggs, and Y. Wang  are with Princeton University, Princeton, NJ 08540, USA (e-mail: smbruno@princeton.edu; mvlungu@gmail.com; page@princeton.edu; staggs@princeton.edu; yuhanw@princeton.edu).}
\thanks{K. T. Crowley is with the Department of Physics, University of California, Berkeley, CA 94720, USA (e-mail: ktc35@berkeley.edu).}
\thanks{S-P. P. Ho and M. Salatino  are with Stanford University and KIPAC, Stanford, CA 94304, USA (e-mail: spho0814@standford.edu; maria5@standford.edu).}
\thanks{J. E. Golec and J. McMahon are with the Department of Physics and the Department of Astronomy and Astrophysics, University of Chicago, Chicago, IL 60637, USA  (e-mail: golecjoe@uchicago.edu; jeff@astro.uchicago.edu).}
\thanks{M. Hasselfield is with Center for the Computational Astrophysics, Flatiron Institute, New York, NY 10010, USA (e-mail: mhasselfield@flatironinstitute.org).}
\thanks{B. J. Koopman is with the Department of Physics, Yale University, New Haven, CT 06520, USA (e-mail: brian.koopman@yale.edu).}
\thanks{S. M. Simon is with the Physics Department, University of Michigan, Ann Arbor, MI 48109, USA (e-mail: smsimon@umich.edu)}
\thanks{E. J. Wollack is with NASA Goddard Space Flight Center, Greenbelt, MD 20771, USA (e-mail: edward.j.wollack@nasa.gov)}
\thanks{Z. Xu is with the Department of Physics and Astronomy,
University of Pennsylvania, Philadelphia, PA 19104, USA (e-mail: zhileixu@sas.upenn.edu) }}

\maketitle

\begin{abstract}
The Advanced Atacama Cosmology Telescope Polarimeter (AdvACT)~\cite{thornton} is an upgrade for the Atacama Cosmology Telescope using Transition Edge Sensor (TES) detector arrays to measure cosmic microwave background (CMB) temperature and polarization anisotropies in multiple frequencies. The low frequency (LF) array was deployed early 2020. It consists of 292 TES bolometers observing in two bands centered at 27 GHz and 39 GHz. At these frequencies, it is sensitive to synchrotron radiation from our galaxy as well as to the CMB, and complements the AdvACT arrays operating at 90, 150 and 230 GHz.  
We present the initial LF array on-site characterization, including the time constant, optical efficiency and array sensitivity.
\end{abstract}


\section{Introduction}

\begin{figure}[H]
  \begin{center}
    \includegraphics[width=0.95\linewidth]{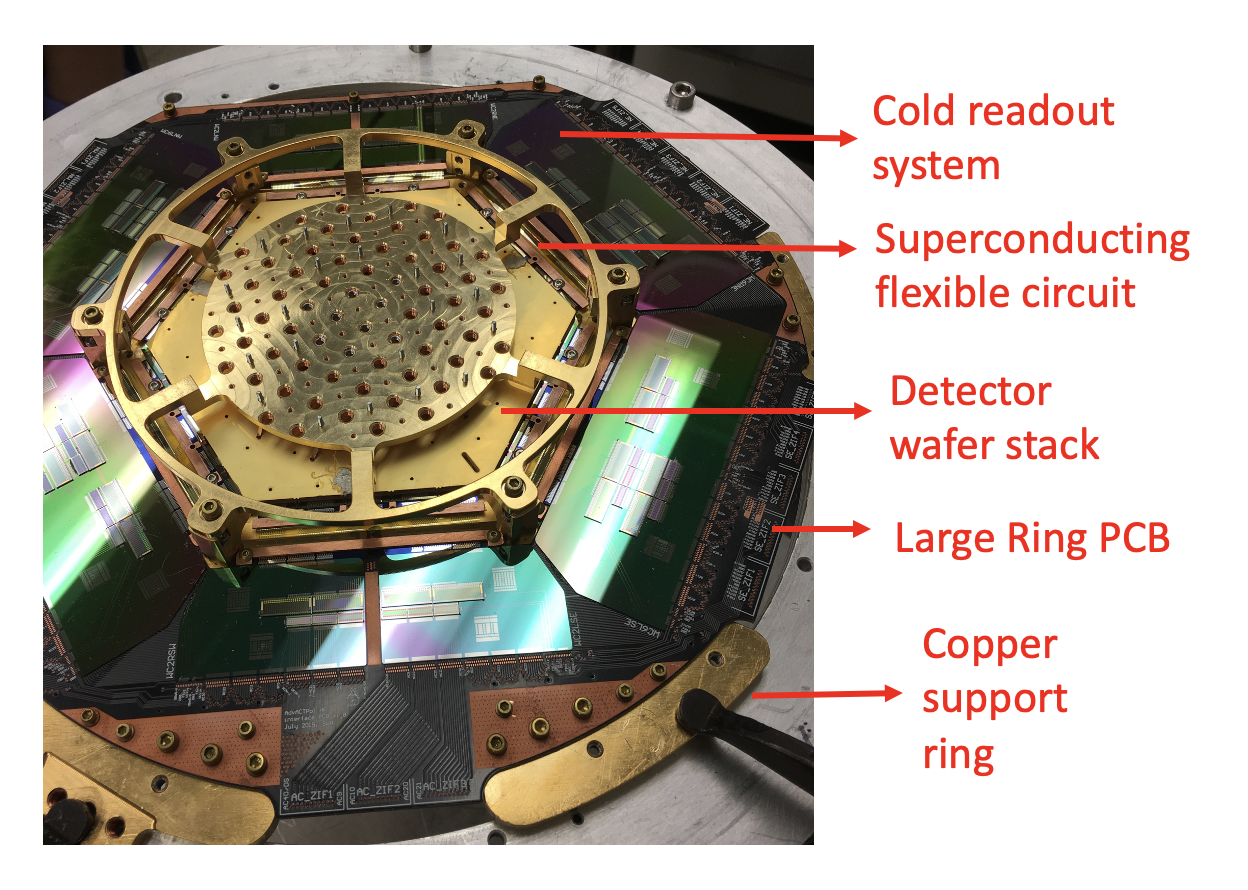}
    \caption{Components of the LF detector array.}
    \label{fig:array}
  \end{center}
\end{figure}

The Advanced Atacama Cosmology Telescope Polarimeter (AdvACT) is an upgraded instrument for the 6~m Atacama Cosmology Telescope using Transition Edge Sensor (TES) bolometer arrays to measure the Cosmic Microwave Background (CMB) temperature and polarization anisotropies with $\sim$arcmin angular resolution, probing the nature of dark matter and dark energy, and the sum of neutrino masses. The receivers include one high frequency (HF)~\cite{ho}, two middle frequency (MF)~\cite{choi} and one low frequency (LF)~\cite{li} multichroic arrays. The HF and MF arrays were deployed and have been observing half the sky. The LF array (shown in Fig.~\ref{fig:array}) was deployed in February 2020 and has been taking measurements of the CMB intensity and polarization anisotropies at 27 and 39 GHz, complementing the HF and MF arrays operating at 90, 150 and 230 GHz and sensitive to synchrotron emission.

\begin{figure}
  \begin{center}
    \includegraphics[width=0.95\linewidth]{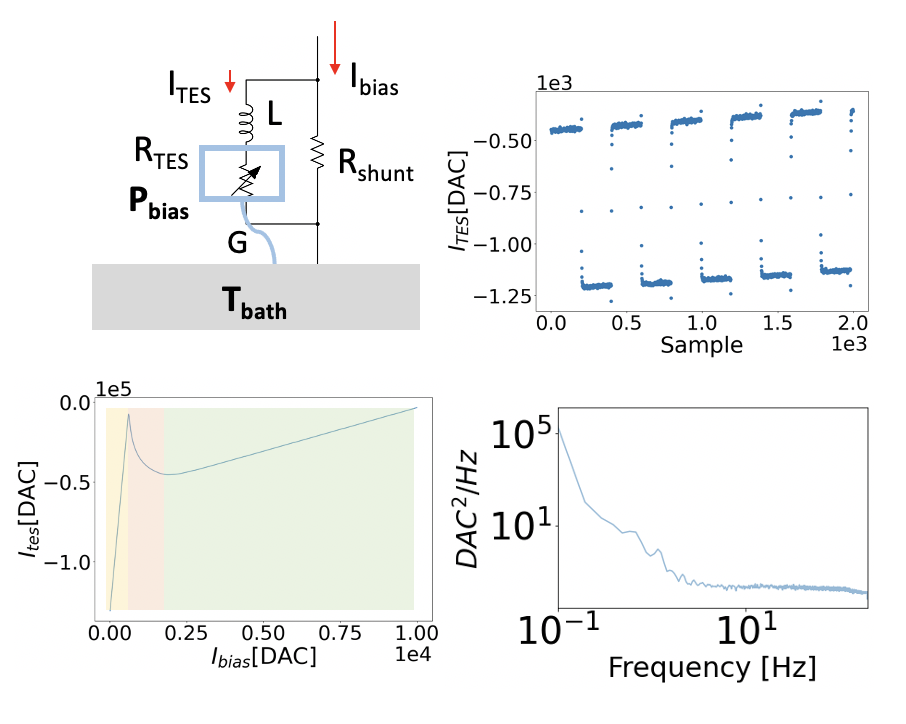}
    \caption{Schematic of TES circuit (Top left). Example of the detector characterization plots, including the current-voltage curve (bottom left), bias step response (top left) and noise spectrum (bottom right). }
    \label{fig:TES_circuit}
  \end{center}
\end{figure}

The LF array consists of 73 horn-coupled pixels, with AlMn TES bolometers that are fabricated on a single 150~mm wafer. Each pixel is composed of an orthomode transducer (OMT) coupled to four TESes, respectively measuring the CMB at two linear polarization and two frequencies. In addition, each pixel has two “dark” TESes that are not coupled to OMT. In total we readout 292 OMT TESes and 98 dark TES. As shown in Fig~\ref{fig:TES_circuit}, Each TES is thermally coupled to bath temperature, $T_{bath}$, and is in parallel with a shunt resistor with $R_{shunt}<<R_{N}$, the TES normal resistance. The TES is operated between the superconducting and normal states, usually at 50\% $R_N$. By sending a fixed current to the shunt resistor, the TES is voltage-biased. This circuit architecture allows the detector to have negative electro-thermal feedback. When incident radiation heats up the TES, the resistance increases, which sequentially decreases the $P_{bias}$, the electrical power dissipated in the TES, which tends to make the detector cool back down to the original state. The signal, in the form of variations of $I_{TES}$, is read out by a time-division multiplexing system (TDM)~\cite{doriese} by inductively coupling the TES circuit to a Superconducting Quantum Interference Device (SQUID).
The readout architecture of the AdvACT array is described in detail in \cite{henderson}.

The LF array provides a unique addition to the ACT's capabilities by enabling high angular resolution measurements across half of the sky in this new waveband.     
We describe the measurements from the field that characterize instrumental parameters including the instrument loading, time constants, array sensitivity and the end-to-end optical efficiency. The characterization and calibration process mainly follow the methods in \cite{ho,choi,grace,hasselfield} 

\section{Telescope Operation and TES Loading}

For each scan, the telescope rotates in azimuth at a fixed elevation, generating time-ordered data (TOD) files that each contains roughly 10~min of data. There are three types of calibrations involved for the LF array. The first type of calibration is based on Jupiter. The planet scans are used to measure the beam profiles and provide information about power-to-temperature conversion factor and time constants. The planet calibration takes place every other day. The second type is measuring the current-voltage curves of the TESes by ramping the detector biases. These IV curves are done at the beginning of each scan, then we reset the bias point so that the TES is kept biased at ~50\% $R_N$. The last type of calibration is the time constant measurement, which will be described in detail in Sec.~\ref{sec:f3db}.

\begin{figure}[H]
  \begin{center}
    \includegraphics[width=0.95\linewidth]{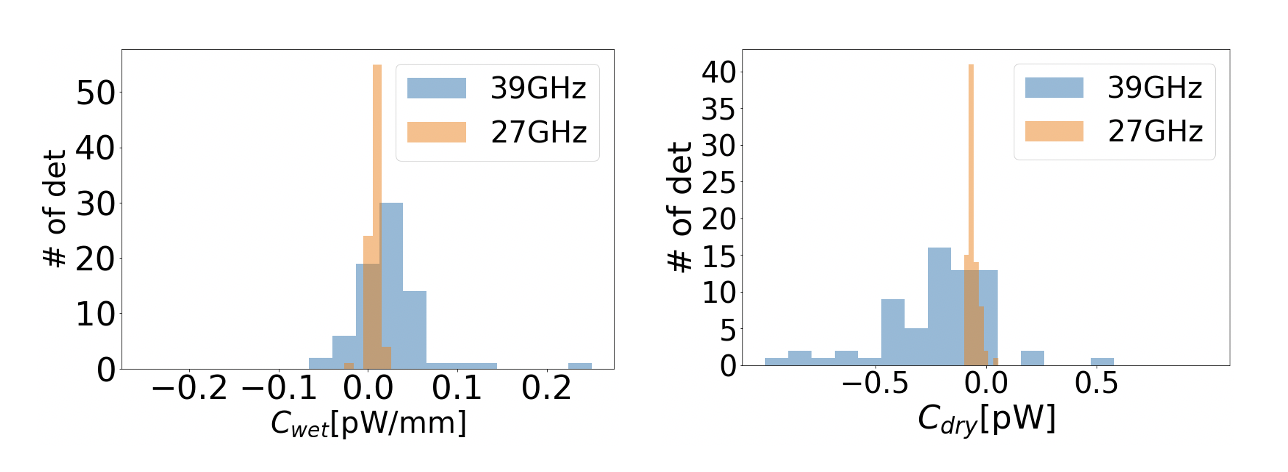}
    \caption{Histograms of $C_{wet}$ (left) and $C_{dry}$ (right) for the LF detectors using the model in Eq~\ref{eq:loading}. We find $C_{wet}$ =0.0077$\pm$0.0053~pW/mm at 27~GHz , and =0.025$\pm$0.030~pW/mm at 39~GHz. After removing $C_{dry}$, we find $C_{wet}$ =0.0066$\pm$0.0057~pW/mm at 27~GHz , and =0.022$\pm$0.029~pW/mm at 39~GHz.}
    \label{fig:atm}
  \end{center}
\end{figure}

The detector noise is dominated by the photon noise, which scales up as the photon power. Therefore minimizing background optical loading is crucial to optimizing the detector sensitivity. The background loading comes from either atmospheric emission, including the emission contributed by the precipitable water vapor (PWV) and other dry components from the atmosphere, or the emission from the optics inside of the telescope. We measure the three components with the model~\cite{grace}:
\begin{equation}\label{eq:loading}
\begin{split}
P_{sat,site}= &\ P_{sat,lab}\\
&\ -\left[P_{instrument}+\frac{C_{wet}PWV}{\sin{\theta_{el}}}+\frac{C_{dry}}{\sin{\theta_{el}}}\right],
\end{split}
\end{equation}
where $P_{sat,site}$ is the bias power needed to saturate the detector to its normal state during operation in the telescope, and $P_{sat,lab}$ is saturation power needed during testing in lab with minimized optical power. The elevation angle is $\theta_{el}$. The $P_{instrument}$, $C_{wet}$, and $C_{dry}$ parameters can then be fitted after a set of measurements with different PWVs and elevation angles. We find $P_{instrument}$ at 39~GHz =1.30$\pm$0.59~pW and $P_{instrument}$ at 27~GHz =0.08$\pm$0.07~pW, and do not observe a significant dependence of loading power on PWV or dry components from the atmosphere (shown in Fig~\ref{fig:atm}). Since fits that included $C_{dry}$ are consistent with zero and sometimes negative, we remove $C_dry$ from the fits to force the fits to follow a more physical solution. The loading power $P_{load}$, which in this case $\sim P_{instrument}$, is widely dispersed and the situation is worse for the 27~GHz detectors. One possible cause is the nonuniformity of the detector optical efficiency, which we describe in more detail in Sec~\ref{sec:sensitivity} and \ref{sec:discussion}.

\section{Time Constants}\label{sec:f3db}
The TES response at high frequency is degraded by the electro-thermal time constant $\tau$, which we model with a single-pole low pass filter with a roll-off at $f_{3dB}=1/(2\pi \tau)$. Therefore, it is crucial to keep track of the time constant and account for the filter effect on the TOD spectra. Each detector has an intrinsic time constant when there is no optical loading and no electrical biasing:
\begin{equation}
\tau=\frac{C}{G}
\end{equation}
where $G$ is the thermal conductance to $T_{bath}$ and $C$ is the TES bolometer capacitance. Both of $C$ and $G$ are determined by the TES bolometer geometry. When the TES is voltage-biased at 50\% $R_N$, because of the negative electro-thermal feedback, the speed of response is greatly boosted up~\cite{irwin}:
\begin{equation}
f_{3dB}=\frac{1}{2\pi}\left(\frac{G}{C}+\frac{\alpha}{1+\beta}\frac{1}{T_{c}C}P_{bias}\right),
\end{equation}
where $T_c$ stands for the critical temperature, and $\alpha$ and $\beta$ are the logarithmic derivative of detector resistance with respect to the detector temperature and current: $\alpha=\frac{T}{R}\frac{dR}{dT}$ and $\beta=\frac{I}{R}\frac{dR}{dI}$. Note that $\alpha$ and $\beta$ depend on the shape of the TES transition curve at the current bias point. Since we are keeping the TES biased at 50\% $R_N$, $\alpha$ and $\beta$ are constant most of the time. In short, $f_{3dB}$ is mostly determined by $P_{bias}$.

\begin{figure}[H]
  \begin{center}
    \includegraphics[width=0.7\linewidth]{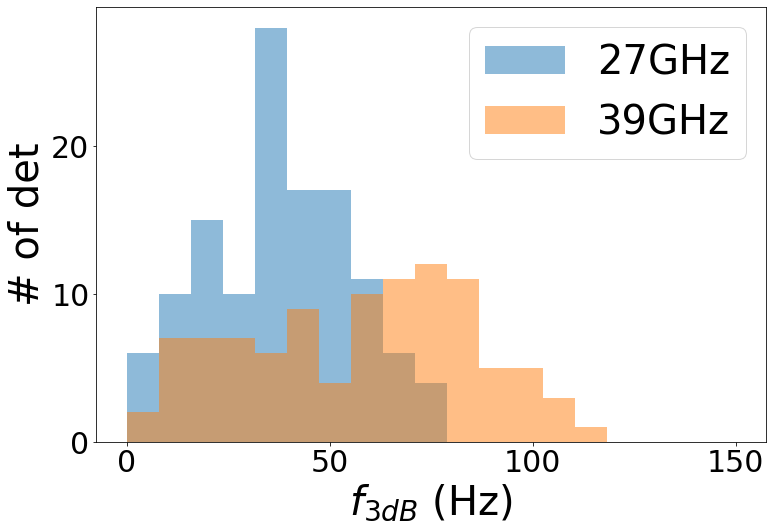}
    \caption{Histogram of $f_{3dB}$ for the LF detectors using the bias step method.}
    \label{fig:f3db}
  \end{center}
\end{figure}

Time constants can be measured by sending square waves to the detector bias lines, so called bias step method. The time constants are found by fitting the detector response at the fallings and risings of square waves to an exponential function: $I \sim e^{t/\tau}$. Fig.~\ref{fig:f3db} shows the histogram of $f_{3dB}$ taken with the bias step method from a set of TOD when PWV was around 1~mm.

\section{Planet Scan and Array sensitivity}\label{sec:sensitivity}

\begin{figure}[H]
  \begin{center}
    \includegraphics[width=0.95\linewidth]{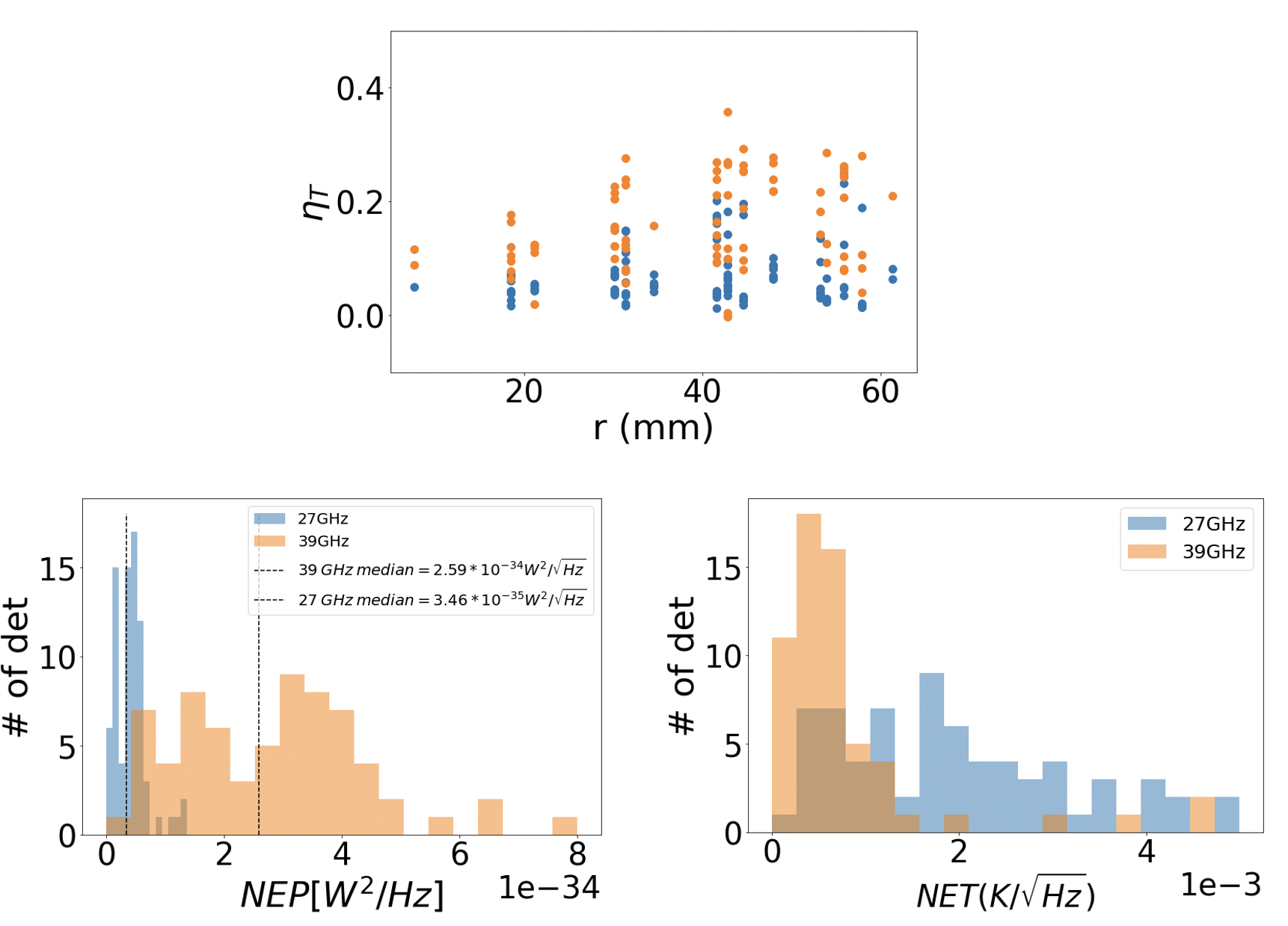}
    \caption{(Top): Plot showing the end-to-end efficiency versus the distance between the pixel and the array center. (Bottom left): Histogram of the NEP for the LF detectors. (Bottom right): Histogram of the calculated NET using the conversion factor.}
    \label{fig:noise}
  \end{center}
\end{figure}

Another crucial measurement we have conducted \emph{in situ} is the noise measurements for individual TESes, from which we estimate the overall array sensitivity. 
The noise spectrum is estimated by the average of the Fourier-transform of the detector TODs (example spectrum shown in Fig.~\ref{fig:TES_circuit}). The white noise is measured as the noise level in a small band at 20~Hz, and it is in the form of Noise Equivalent Power (NEP), with units $\rm pW/\sqrt{\rm Hz}$. The left side of Fig.~\ref{fig:noise} shows the histogram of NEP for PWV$=\sim$1~mm.

We use measurements of Jupiter to estimate the conversion factor between power fluctuations and CMB temperature fluctuations. Jupiter's brightness temperature is well known. The solid angle of Jupiter is considerably smaller than the detector beam so it can be considered as a point source and used to measure the detector beam. The antenna temperature $T_a$ is diluted by the detector beam~\cite{choi}:
\begin{equation}
T_a=\frac{\Omega_p}{\Omega_b}T_p,
\end{equation}
where $\Omega_p$ and $\Omega_b$ are the solid angles of the planet and detector beam, and $T_p$ is the brightness temperature of the planet. The $T_p$ values are taken from WMAP~\cite{WMAP}. The maps of Jupiter are generated using the code \emph{moby2}\footnote{ https://github.com/ACTCollaboration/moby2}~\cite{hasselfield}. The code is also used for fitting the map to compute $\Omega_b$, which contains the Gaussian main lobe and ``wing" part that decays as $1/\theta^3$, where $\theta$ is the angular distance to the center of the beam. Under the assumption that the spectrum of the brightness temperature is relatively flat around the detector bandwidth $\Delta \nu$, the optical power measured by the detector becomes:
\begin{equation}\label{eq:validataion:insitu:eta}
\begin{split}
P_{\gamma}= 
&\ \frac{1}{2}\int\eta_{T}\frac{c^2}{\nu ^2}\frac{\Omega_p}{\Omega_b}B(\nu,T)d\nu\\
&\ \sim \eta_T\int k_{B} \frac{\Omega_p}{\Omega_b}T_{p}d\nu\\
&\ \sim k_{B}\eta_{T} \frac{\Omega_p}{\Omega_b}T_{p}\Delta \nu 
\end{split}
\end{equation}
where $\eta_{T}$ is the end-to-end efficiency of the telescope, and $B(\nu,T)$ is Planck's law. The $P_{\gamma}$ is measured as the peak amplitude (in TES bias power) from Jupiter for each detector during the planet scan. The right side of Fig.~\ref{fig:noise} shows the result of $\eta_{T}$. Since $\eta_T$ includes the coupling between the feedhorn beam and the Lyot stop, it tends to decrease and scatter more as the detector is further from the array center. 

The conversion factor can be calculated from the planet observation by:
\begin{equation}\label{eq:validation:conversion_factor}
\frac{\delta T_{CMB}}{\delta P} = \frac{\Omega_p T_p}{\Omega_b P_{\gamma}}.
\end{equation}
With this calibration factor, we then convert NEP to Noise Equivalent Temperature (NET) in units  $\rm \mu \rm K/\sqrt{\rm Hz}$. The array overall NET for each frequency band is obtained by inverse-variance-weighting:

\begin{equation}\label{eq:array_sensitivity}
NET_{array}^2= \frac{1}{\sum_{i}\frac{1}{NET_{i}}^2},
\end{equation}
where $NET_i$ is NET of each detector belonging to the frequency band. The final estimated results of $NET_{array}$ are 36~\text{$\rm \mu K\sqrt{\rm s}$} at 39~GHz and 72~\text{$\rm \mu K\sqrt{\rm s}$} at 27~GHz relative to the CMB.

\section{Discussion}\label{sec:discussion}

The target values of the array sensitivity are 35~\text{$\rm \mu K\sqrt{\rm s}$} at 27~GHz and 32~\text{$\rm \mu K\sqrt{\rm s}$} at 39~GHz. The estimated $NET_{array}$ at 39~GHz is close to the target, while $NET_{array}$ at 27~GHz is significantly higher than expected. The main reason for the higher NET level is the nonuniformity of the detector optical efficiencies. As the distance between the detector and the array center increases, the optical efficiency becomes more widely dispersed, and the detectors in a same pixel at different polarizations tend to have more distinct optical responses. We have experienced the same issue on the HF array~\cite{ho}, but for the MF arrays, the amount of this dependence on polarizations is reduced. Further research is needed to investigate the source of this issue.

\section{Conclusion}

The LF array received its first light in February of 2020. We report the measured parameters (summarized in Table~\ref{tab1}) including time constants and the array sensitivity. The method to characterize it will shine a light on the low frequency detector arrays in future generation of CMB instruments.

\begin{table}[H]
\caption{Parameters of the LF detectors.}
\begin{center}
\begin{tabular}{|c|c|c|}
\hline
\textbf{Freq}& \textbf{27GHz}& \textbf{39GHz}    \\
\hline
\textbf{Num}& 123 & 97    \\
\hline
\boldmath{$\Omega_b$}\textbf{[nsr]}& 4165$\pm$317 & 3064$\pm$1204    \\
\hline
\textbf{FWHM[arcmin]}& 6.8$\pm$0.5 & 5.7$\pm$0.7    \\
\hline
\boldmath{$f_{3dB}$}\textbf{[Hz]}& 48.1$\pm$27.2 & 64.4$\pm$24.5    \\
\hline
\textbf{NET[}\boldmath{$\rm \mu K\sqrt{\rm s}$}\textbf{]}& 72 & 36    \\
\hline
\end{tabular}
\label{tab1}
\end{center}
\end{table}




\begin{thebibliography}{00}

\bibitem{thornton} R. J. Thorton \emph{et al}., "The Atacama Cosmology Telescope: The polarization-sensitive ACTPol instrument," \emph{Astrophys. J. Suppl. Ser.}, vol. 227, no. 2, 2016, Art. no. 21. DOI:10.3847/1538-4365/227/2/21.

\bibitem{ho} S-P. P. Ho \emph{et al}., "Highly uniform 150 mm diameter multichroic polarimeter array deployed for CMB detection," \emph{Proc. SPIE}, vol. 9914, 2016, Art. no. 991418. DOI:10.1117/12.2233113

\bibitem{choi} S. K. Choi \emph{et al}., "Characterization of the Mid-Frequency Arrays for Advanced ACTPol," \emph{J. Low Temp. Phys}., vol. 193, pp. 267–275, Jun. 2018. DOI:10.1007/s10909-018-1982-4

\bibitem{li} Y. Li \emph{et al}. "Performance of the advanced ACTPol low frequency array", \emph{Proc. SPIE}, vol. 10708, 2018, Art. no. 107080A. DOI: 10.1117/12.2313942.


\bibitem{henderson}
S. Henderson \emph{et al}. “Advanced ACTPol cryogenic detector arrays and readout,” \emph{J. Low Temp}. Phys., vol. 184, pp. 772–779, 2016. DOI: 10.1007/s10909-016-1575-z.

\bibitem{doriese}W. B. Doriese \emph{et al}., “Developments in time-division multiplexing of X-ray transition-edge sensors,” \emph{J. Low Temp. Phys.}, vol. 184, no. 1–2, pp. 389–395, 2016. DOI:10.1007/s10909-016-1575-z.

\bibitem{grace}E. Grace \emph{et al}. "ACTPol: on-sky performance and characterization", \emph{Proc. SPIE}, vol. 9153, 2014, Art. no. 915310. DOI: 10.1117/12.2057243.

\bibitem{hasselfield}M. Hasselfield \emph{et al}. "The Atacama Cosmology Telescope: Beam Measurements and the Microwave Brightness Temperatures of Uranus and Saturn", \emph{ApJS}, vol. 209, 2013, Art. no. 1. DOI: 10.1088/0067-0049/209/1/17.

\bibitem{irwin} K. D. Irwin and G. C. Hilton, “Transition-edge sensors,” in \emph{Cryogenic Particle Detection} (Topics in Applied Physics series 99). New York, NY, USA: Springer, 2005, pp. 63–150.

\bibitem{WMAP} 
J. L. Weiland \emph{et al}. "Seven-Year Wilkinson Microwave Anisotropy Probe (WMAP) Observations: Planets and Celestial Calibration Sources", \emph{ApJS}, vol. 192, 2011, Art. no. 2. DOI: 10.1088/0067-0049/192/2/19/meta.



\end{thebibliography}
\end{document}